\documentclass[preprint]{aastex}

\usepackage{amsmath}
\usepackage{graphicx}
\usepackage{hyperref}
\usepackage{subfig}
\usepackage[final]{pdfpages}
\usepackage{natbib}
\newcommand{\snr}{{\raise0.7ex\hbox{${\rm{S}}$} \!\mathord{\left/
 {\vphantom {{\rm{S}} {\rm{N}}}}\right.\kern-\nulldelimiterspace}
\!\lower0.7ex\hbox{${\rm{N}}$}}
}

\begin{document}

\author{Andrew P.V. Siemion\altaffilmark{1}, Geoffrey C. Bower\altaffilmark{1}, Griffin Foster\altaffilmark{2}, Peter L. McMahon\altaffilmark{3}, Mark I. Wagner\altaffilmark{1}, Dan Werthimer\altaffilmark{1}, Don Backer\altaffilmark{1}, Jim Cordes\altaffilmark{4}, Joeri van Leeuwen\altaffilmark{5}}

\email{siemion@berkeley.edu}


\altaffiltext{1}{University of California, Berkeley}
\altaffiltext{2}{Oxford University}
\altaffiltext{3}{Stanford University}
\altaffiltext{4}{Cornell University}
\altaffiltext{4}{Netherlands Institute for Radio Astronomy (ASTRON)}

\pagenumbering{roman}
\title{The Allen Telescope Array Fly's Eye Survey for Fast Radio Transients}
\maketitle
\setcounter{page}{1}
\pagenumbering{arabic}

\section*{Abstract}
The relatively unexplored fast radio transient parameter space is known to be home
to a variety of interesting sources, including pulsars, pulsar giant pulses and non-thermal emission from planetary magnetospheres. In addition, a variety of hypothesized but as-yet-unobserved phenomena, such as primordial black hole evaporation and prompt emission associated with coalescing massive objects have been suggested. 
The 2007 announcement by Lorimer et al. of the detection of a bright (30 Jy) radio pulse that was inferred to be of extragalactic origin and the subsequent consternation have demonstrated both the potential utility of bright radio pulses as probes of the interstellar medium and intergalactic medium, as well as the need for wide-field surveys characterizing the fast-transient parameter space.  Here we present results from the 450 hour, 150 deg$^2$ Fly's Eye survey for bright dispersed radio pulses at the Allen Telescope Array (ATA).  The Fly's Eye spectrometer produces 128 channel power spectra over a 209 MHz bandwidth, centered at 1430 MHz, on 44 independent signals paths originating with 30 independent ATA antennas.  Data were dedispersed between 0 and 2000 pc cm$^{-3}$ and searched for pulses with dispersion measures greater than 50 pc cm$^{-3}$ between 625 $\mu$s and 5 s in duration.  No pulses were detected in the survey, implying a limiting rate of less than 2 sky$^{-1}$ hour$^{-1}$ for 10 millisecond duration pulses having apparent energy densities greater than 440 kJy $\mu$s, or  mean flux densities greater than 44 Jy.  Here we present details of the instrument, experiment and observations, including a discussion of our results in light of other single pulse searches. 

\section{Introduction}
The last decade has seen an explosion of interest in time domain radio astronomy.  Driven by new wide field survey instruments, multi-beam receivers and computing advances, exploration of this regime presents opportunities to shed new light on known phenomena and perhaps reveal previously unseen processes as well \citep{Cordes:2007p12089}.  Investigations in time domain radio astronomy can be conveniently divided into two categories: those that deal withâ slow transients, events lasting from hours to years, and those that deal withâ fast transients, shorter duration events lasting from nanoseconds to seconds.  Slow transients may originate from a diverse set of phenomena, ranging from the radio counterparts to relatively nearby supernovae or explosive events observed at very high redshift to variation in flux from radio jets associated with accretion disks, as discussed at length in \cite{Bower:2007p12091}, \cite{Ofek:2010p12092}, \cite{Bower:2011p10634} and \cite{Becker:2010p10760}.  Fast radio transient phenomena have been observed to arise from only two fundamental physical origins: electrostatic discharge in planetary atmospheres, e.g. \cite{Zarka:2008p1921}, and non-thermal magnetospheric emission from planets e.g. \cite{Zarka:2007p10606} or stars e.g. \cite{Lorimer:2005p6913, Hallinan:2008p10699}.  By far the most commonly explored source of fast radio transient emission are pulsars, including their relatively more intermittent flavors: ``rotating radio transients'' (RRATs), nulling pulsars and those producing giant pulses.  A variety of as-yet-unobserved sources of fast transient emission have been proffered, among the more commonly cited are evaporating primordial black holes \citep{Rees:1977p3912} and compact object coalescence \citep{Hansen:2001p3687}.  Other more speculative sources include emission from cosmic strings \citep{Vachaspati:2008p10774} and beacons from extraterrestrial intelligence \citep{Benford:2008p614}. 

The fast/slow division arises primarily from the different detection techniques employed in each case.  Slow transients are identified at radio wavelengths using essentially the same techniques as used at optical wavelengths; images (usually synthesized from interferometric antenna arrays) from multiple epochs are differenced and thresholded, with the period between epochs driven by the parameters of a given experiment.  Fast transients are generally identified via a method well established from decades of pulsar searches: high time resolution power spectra are corrected for the dispersive effects of the interstellar medium, dedispersed, assuming different integrated column densities of free electrons, dispersion measuresâ or DMs, then collapsed to a time series and thresholded for impulsive events.  In the case of periodicity searches a Fourier transform or fast folding algorithm is also applied to the time series.  Although the existence of nulling pulsars has been known for some time \citep{Backer:1970p10501}, only fairly recently have pulsar searches targeted them specifically.  Following the \cite{McLaughlin:2006p6248} discovery of the extremely intermittent class of neutron stars dubbed RRATs, several archival data sets were reanalyzed including specific single pulse search algorithms in addition to periodicity searches.  Although nearly all RRATs and nulling pulsars have been shown to possess an underlying normal periodicity, the duty cycle of emission can render some pulsars undetectable in a folded profile or Fourier transformed power spectra.  For very low duty cycle objects or those with extreme variation in pulse flux, single pulse searches can be much more effective than periodicity searches and sometimes the only viable method of detection.

In late 2007, a particularly perplexing radio transient was observed in an archival search of 1.4 GHz pulsar survey data from the Parkes multibeam receiver -- the so-called â``Lorimer Burstâ''  \citep{Lorimer:2007p5652}.  This very bright single pulse was detected at a very high signal-to-noise ratio, with a peak flux nearly 100 times the search threshold, in a pointing a few degrees south of the Small Magellanic Cloud (SMC).  The pulse clearly exhibited the quadratic chirp expected from an astrophysical pulse modified by a cold plasma dispersion relation, with an inferred DM of 375 pc cm$^{-3}$.  Even liberal models for the galactic and SMC contribution to the total implied electron column density could account for only a fraction of the DM measured.  Assuming the rest of the dispersion was due to a Milky Way-like host interstellar medium (ISM) contribution and traversal of the much more rarified intergalactic medium (IGM), the lower limit on the distance to the source was calculated to be $\sim$600 Mpc.  Suffice it to say, the implied energy release of $\sim10^{40}$ ergs presented a challenge for astrophysical theory, and motivated wide ranging speculation of possible origins.  Several subsequent searches \citep{Keane:2010p5855, Deneva:2009p4149} did not detect any similar events, implying that such events must be exceedingly rare.  \cite{BurkeSpolaor:2010p5814} presented the detection of several additional impulsive events in Parkes survey data with similar dispersive chirps to that seen in the Lorimer Burst but exhibiting clear indications of terrestrial origins.  While the \cite{BurkeSpolaor:2010p5814} events showed an approximately quadratic frequency evolution, as would be expected for an astrophysical event, the detection of the events in multiple receiver beams simultaneously clearly points to a terrestrial source and the irregularity of received flux across the observing band and large pulse width differentiate them markedly from the Lorimer Burst.  Recently, another possibly extragalactic burst was discovered in additional re-analysis of Parkes survey data \citep{Keane:2011p10822}, lending some support for the existence of a bonafide population of very bright extragalactic fast transient sources.  Regardless of the source of such bursts, a population of extragalactic objects or events producing extraordinarily energetic radio pulses would provide an invaluable probe of the ionized IGM.    

Here we present a search using the 42-dish Allen Telescope Array for bright dispersed radio pulses, with specific attention paid to those of possible extragalactic origin.  Section \ref{sec:inst} describes the digital spectrometer developed for this experiment, installation, verification and calibration procedures, Sections \ref{sec:obs} and \ref{sec:ana} detail observations and analyses and Section \ref{sec:discussion} presents our results and interpretation.

\section{Instrument and Installation}
\label{sec:inst}
The single pulse search described here used the Allen Telescope Array (ATA) \citep{Welch:2009p932} in an unconventional non-interferometric mode.  Rather than pointing all 42 dishes in the same direction, each dish was pointed at a unique position, similar to a ``fly's eye.''   Such a mode yields a dramatically increased field of view at the expense of sensitivity, well matched to detecting bright, rare events.  The primary half-power beam width (HPBW) of the ATA is approximately 2.5 deg at 1.4 GHz, yielding a potential field of view of more than 200 deg$^2$.  

In this experiment, each antenna signal path was processed independently using a purpose-built digital spectrometer.  This device, dubbed the Fly's Eye Spectrometer, was constructed using the modular instrumentation infrastructure developed by the Center for Astronomy Signal Processing and Electronics Research \citep{Werthimer:2011p9714}.  The full system consists of eleven field programmable gate-array (FPGA)-based `iBOB' computing boards, each equipped with two 1024 Msample/sec `iADC' analog-to-digital converter cards.  Each ADC board digitizes two independent single-polarization signal paths at 838.8608 Msamples/sec. The Nyquist sampled band is digitally down converted and decimated to a bandwidth of 209.7152 MHz within the FPGA and channelized using a 2$^7$ channel complex biplex-pipelined polyphase filterbank.  Power spectra are detected and accumulated for 625 $\mu$s, packetized into Ethernet UDP packets on each of the eleven FPGA boards and transmitted to a single Linux PC via an Ethernet switch.  The payload of each UDP packet contains a 21 byte header, which specifies a board ID, accumulation number and any error conditions, followed by 512 bytes of spectral data (unsigned byte power measurements $\times$ 128 frequency channels $\times$ 4 inputs).  The entire Fly's Eye digital back end is described in detail in \cite{McMahon:2008p6817} and \cite{Siemion:2010p6845}.  Large portions of the hardware design and software are open source and freely available at http://casper.berkeley.edu.

During the course of analyzing our initial observations, a subtle timing error was discovered in the accumulation counter produced by the Fly's Eye hardware.  This error was initially attributed to a flaw in the digital design for the instrument involving the 1 pulse per second synchronization logic.  As a result, we disconnected the 1 pulse-per-second input signal from each iBOB board prior to beginning our observation campaign.  It was later discovered that the principal cause of the timing error was not the flaw in the digital design, but rather that the sampling clock was not properly locked to the observatory reference.  By comparing the unix time stamps applied to each spectra by our data collection computer with the hardware counter applied by the Fly's Eye spectrometer, we determined that the sampling clock frequency was only absolute to about 1 part in $10^{4}$.  This level of error translates to a center RF frequency and bandwidth ambiguity of about 84 kHz.  Because the Fly's Eye spectrometer derives integration time from counting the sampling clock, the unlocked clock synthesizer also imposes an ambiguity in integration time of about 60 nanoseconds.  In total, these effects correspond to an additional $\sim$0.03\% temporal smearing at 1430 MHz.  Although a small amount of additional temporal smearing has only a small impact on single pulse searches, it renders simple barycentering and folding of known pulsars impossible.  We accounted for this effect during operability determination by `searching' for a known pulsar rather than directly folding on its known ephemeris.   

\begin{figure}[htb]
\includegraphics[width=0.4\textwidth]{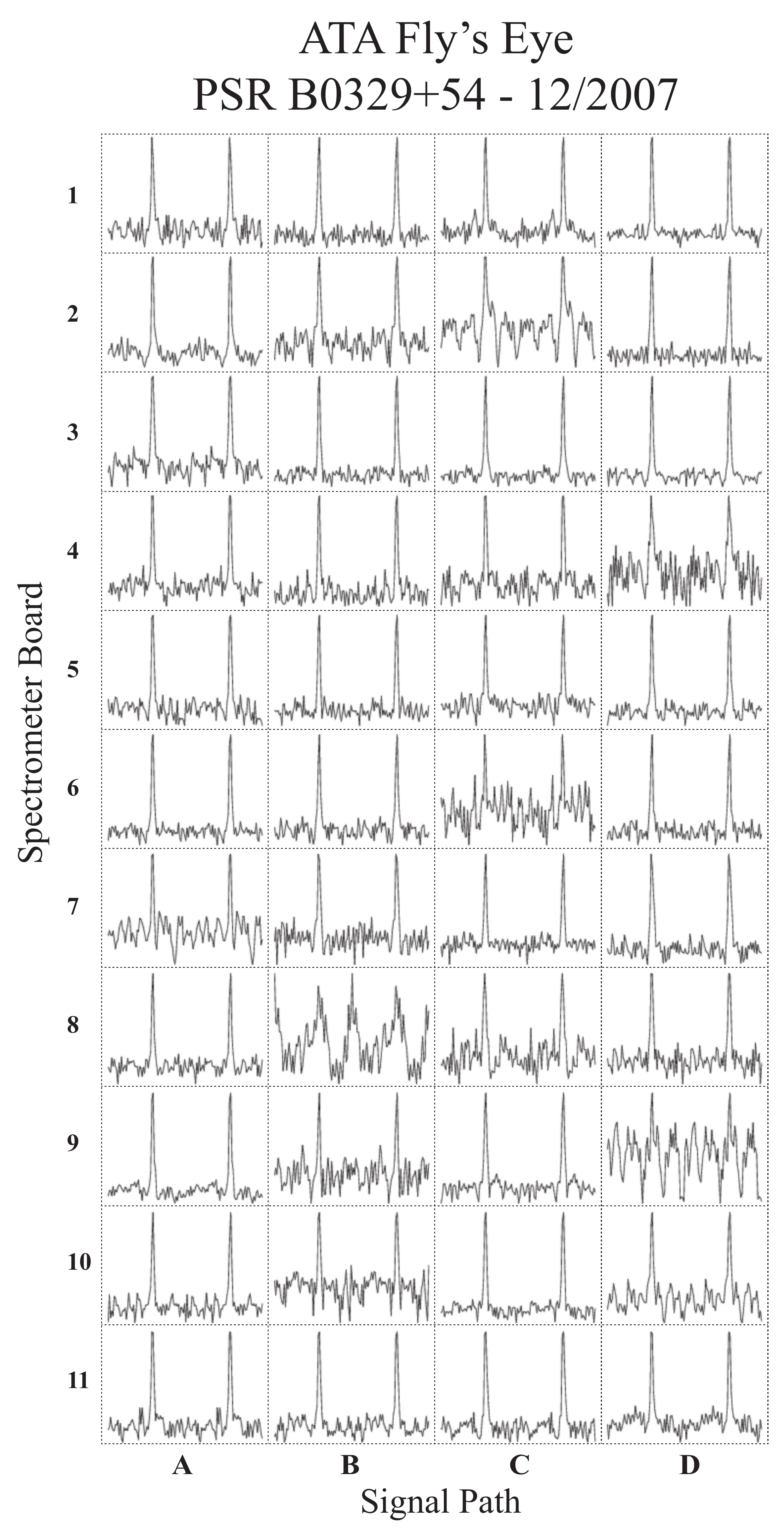}
\caption{
Folded pulse profiles of PSR B0329$+$54 as observed in individual Fly's Eye Spectrometer inputs for a 1290s integration at 1430 MHz.  Two full turns plotted for clarity
\label{fig:0329detections}
}
\end{figure}
\begin{figure}[htb]
\includegraphics[width=0.6\textwidth]{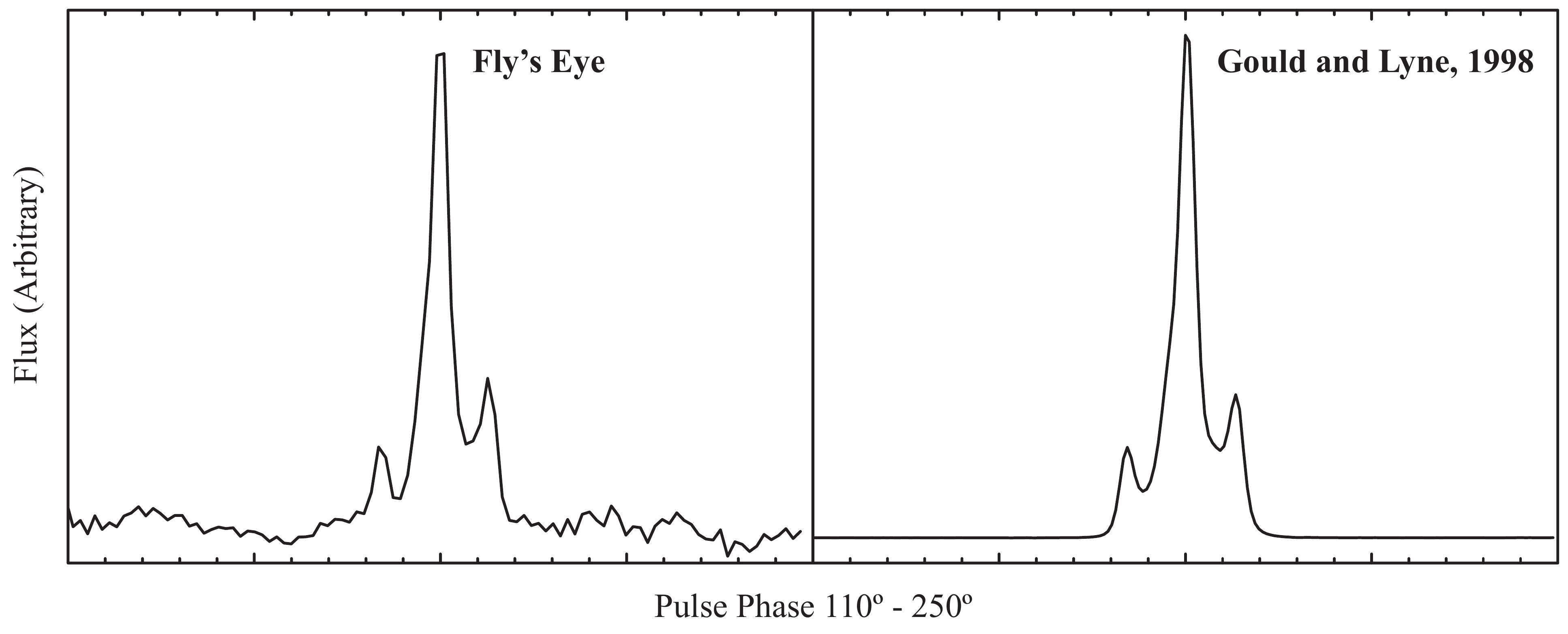}
\caption{
PSR B0329$+$54 folded pulse profile as detected by the incoherent sum of all 44 inputs to the Fly's Eye Spectrometer (shown in \ref{fig:0329detections}) for a 1290s integration at 1430 MHz (left) and a reference profile from \cite{Gould:1998p9871} taken at 1408 MHz. 
\label{fig:0329profile}
}
\end{figure}

The Fly's Eye Spectrometer system was installed at the Allen Telescope Array in December of 2007 and connected to the 44 antenna-polarizations exhibiting the lowest system temperature.   Galactic HI detection along several lines of sight served as a coarse operability check.  To further test the Fly's Eye Spectrometer, and most importantly to test our ability to detect bright dispersed pulses, we observed several bright pulsars including PSR B0329$+$54, the brightest known pulsar in the northern celestial sphere in terms of mean flux and also PSR B0531$+$21 (the Crab pulsar) - a canonical source of bright dispersed radio pulses \citep{Bhat:2008p10164, Cordes:2004p7058}.  Figure \ref{fig:0329detections} shows folded pulse profiles for a 1290s B0329$+$54 integration as observed in each of the 44 Fly's Eye inputs.  All inputs but one (8B) show some detection, with only 4 others (9D, 6C, 4D, 2C) showing significantly degraded signal-to-noise.  Figure \ref{fig:0329profile} shows an unweighted incoherent sum of the same data plotted alongside a reference profile.    

The signal-to-noise ratio of a detected pulsar pulse profile can be calculated as \citep{Cordes:1998p8550}:
\begin{equation}
{\raise0.7ex\hbox{$S$} \!\mathord{\left/
 {\vphantom {S N}}\right.\kern-\nulldelimiterspace}
\!\lower0.7ex\hbox{$N$}} = \frac{{S_{\rm mean} }}{{S_{\rm sys} }}\sqrt{n_{\rm p} t_{\rm int} \Delta f N_{\rm h}}
\end{equation}
where ${\raise0.7ex\hbox{$S$} \!\mathord{\left/{\vphantom {S N}}\right.\kern-\nulldelimiterspace}\!\lower0.7ex\hbox{$N$}}$ is the detected signal
to noise ratio, $S_{\rm{mean}}$ is the mean flux density of the pulsar, $S_{\rm{sys}}$ is the system equivalent flux density (SEFD) of the observing system, $n_{\rm p}$ the number of polarizations summed, $t_{\rm int}$ the integration time, $\Delta f$ the bandwidth observed and $N_{\rm h}$ is the number of harmonics used in a harmonic sum, which depends on the pulse period $P$ and pulse width $W$ as $N_{\rm h} \approx P/W$.  For the case of multiple antennas sampled synchronously and detected independently, $n_{\rm{\rm p}}$ is equal to the total number of antenna-polarization signals incoherently summed, $n_{\rm{ant-pol}}$.  Applying this equation to the summed profile shown in Figure \ref{fig:0329profile} with $n_{\rm{ant-pol}}=44$ and the expected mean SEFD for individual ATA antennas ($\sim$10 kJy), we infer a flux density for B0329$+$54 of $S_{\rm{mean}} \approx 100$ mJy.  This value is in reasonable agreement with a mean value extrapolated from \cite{Manchester:2005p6948}, $S_{\rm{mean}}=190$ mJy, especially considering that the pulsar B0329$+$54 has a variable mean observed flux density of a factor of $\sim$3 at 1400 MHz \citep{Wang:2005p4226, Liu:2006p4193}.  PSR B0329$+$54 is thus a poor flux calibrator, but it is never-the-less a useful diagnostic source for a low sensitivity pulse search.



\begin{figure}[htb]
\includegraphics[width=0.6\textwidth]{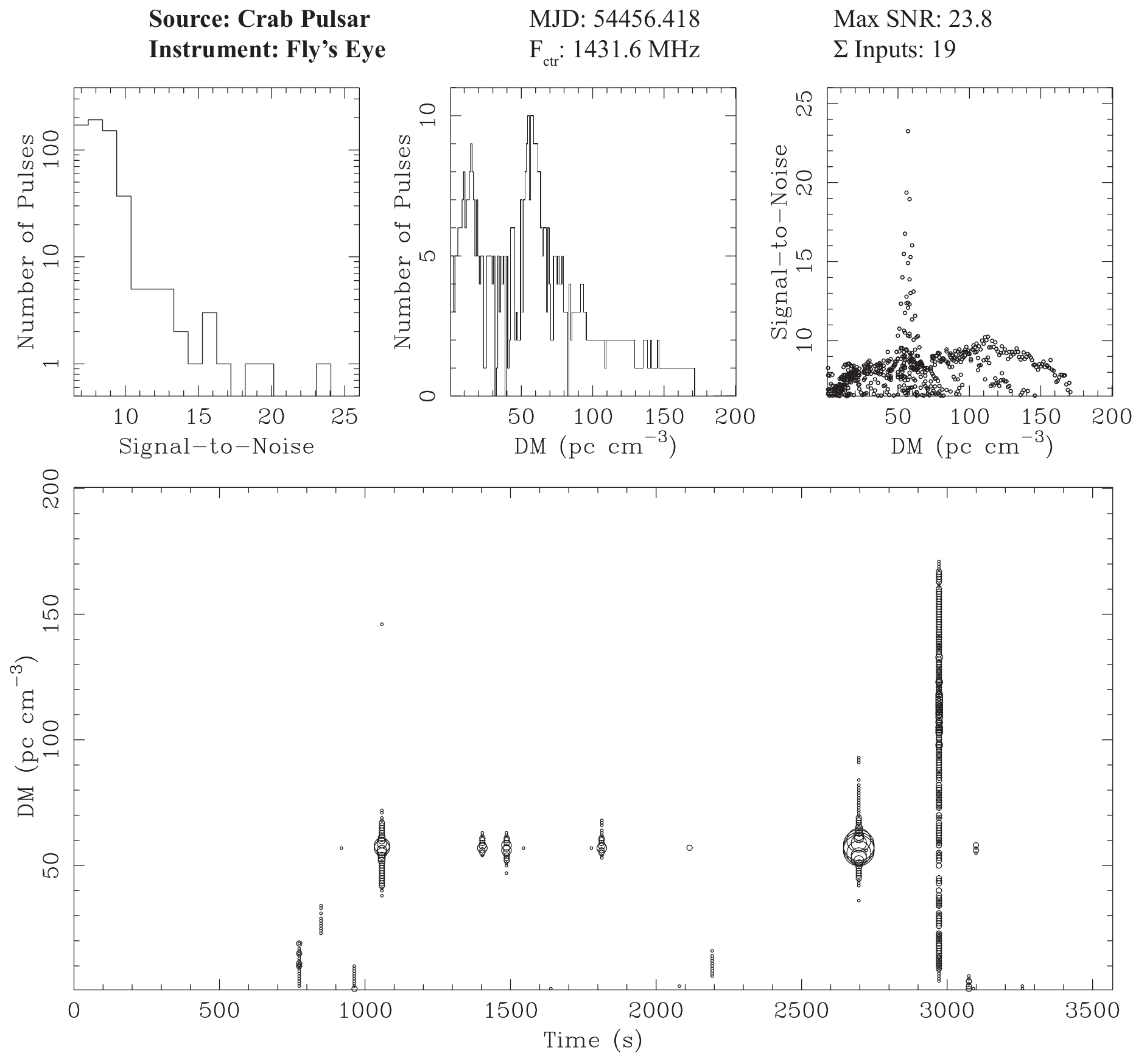}
\caption{
A standard single pulse detection plot for a $\sim$50 minute observation of the Crab pulsar after summing 19 Fly's Eye inputs.  Several giant pulses are apparent at a DM of 56.8 pc cm$^{-3}$.  Other features are radio frequency interference.  From top, left to right, the panels show a histogram of detection signal-to-noise ratio (SNR) for pulses $>$ 5$\sigma$, a histogram of detection dispersion measure, detection dispersion measure against detection SNR and detection time vs. dispersion measure with detection signal to noise indicated by plot point radius.  This plot produced using PRESTO \citep{Ransom:2001p10819}.    
\label{fig:crabgp}
}
\end{figure}

Observation of giant pulses (GPs) from the Crab pulsar provided the final end-to-end test of the Fly's Eye observing system. Crab GPs are known to follow a power law brightness distribution, often parameterized in terms of a cumulative probability distribution $P(E_{\rm i} > E_0 ) = KE_0 ^\alpha$, where $P$ gives the probability of a pulse having a pulse area $E_{\rm i}$ greater than $E_0$.  Here pulse area is defined as $E_{\rm{i}}=S_{\textrm{i}}W_{\textrm{i}}$, where $S_{\textrm{i}}$ is the mean intrinsic flux density of the pulse over an intrinsic time $W_{\textrm{i}}$.  While other authors have referred to the quantity $E_{\rm{i}}$ as ``energy'', in later portions of this work we will use the slightly more accurate ``energy density.''  At 1300 MHz, \cite{Bhat:2008p10164} gives $\alpha\sim$ -1.9 for energy densities greater than 10 kJy $\mu$s, with $K = 4.7 \times 10^{-2}$  and $E_0$ in kJy $\mu$s.  Rearranging equation 3 in \cite{Deneva:2009p4149} gives an expression for the minimum detectable energy density, 
\begin{equation}
E_{\rm{i}}  = \frac{m{S_{{\rm{sys}}} \sqrt{W} }}{{\sqrt {n_{\rm p} \Delta f} }} 
\end{equation}
$W$ is the observed pulse width, usually taken to be the quadrature sum of the various sources of broadening, both astrophysical and instrumental and $m$ the signal-to-noise threshold.  For a pulse with an observed width limited by our digital hardware and assuming a single polarization SEFD of $\sim$10 kJy, a characteristic $m=5\sigma$ minimum detectable energy density for the incoherent sum of 19 inputs is 20 kJy $\mu$s or an mean flux density of $\sim$32 Jy for a 625 $\mu$s pulse.  Based on the expected Crab GP distribution, we should observe a pulse with $E_{\rm{i}} > 20$ kJy $\mu$s about 18 times per hour on average.  Figure \ref{fig:crabgp} shows the detection of 10 bright GPs at the expected DM of 56.8 pc cm$^{-3}$ for the Crab pulsar in a 50 minute unweighted incoherently summed observation using the 19 best performing FE inputs, as determined from Figure \ref{fig:0329detections}.

    
\begin{figure}[htb]
\includegraphics[width=0.4\textwidth]{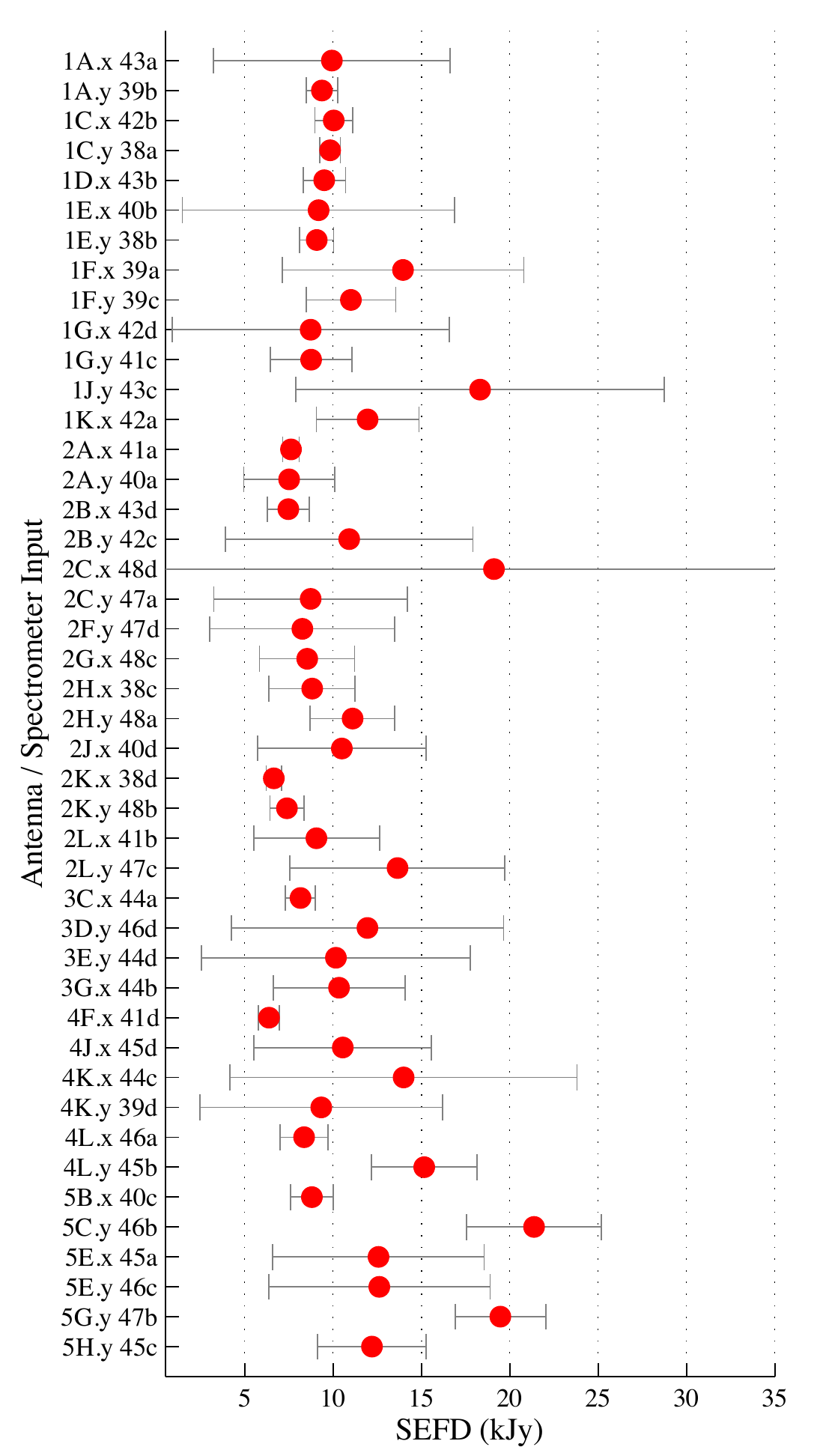}
\caption{
Mean antenna-polarization performance for each input to the Fly's Eye Spectrometer, as determined by interferometric observation of standard calibrators interleaved with Fly's Eye observations between 02/2008 and 05/2008.  Errors are $\pm$ 1$\sigma$.
\label{fig:sefd}
}
\end{figure}

The set of 44 antenna inputs ultimately used for the FE observing campaign originated with 30 independent antennas, 14 of which included both X and Y polarizations.  Figure \ref{fig:sefd} shows antenna performance, described by the system equivalent flux-density (SEFD), for each of the 44 inputs chosen for the Fly's Eye observing campaign.  The values given here were determined by interferometric observation of standard calibrators interleaved with Fly's Eye observation.  Details of the Fly's Eye instrument parameters are given in Table \ref{tab:feparam}.

\begin{table}[h!]
\begin{minipage}{3.2in}
\footnotesize
\caption{Fly's Eye Instrument Parameters}
\centering
\vspace{0.1in}
\begin{tabular}{lcl}
\hline\hline\\[-0.10in]
\textbf{Center Frequency} & $\nu_o$ & 1430 MHz \\
\hspace{0.2in}\\[-0.1in]
\textbf{Number of Channels} & $N_{\textrm{chan}}$ & 128 \\
\hspace{0.2in}\\[-0.1in]
\textbf{Channel Width} & $\Delta\nu_i$ & 1.6 MHz \\
\hspace{0.2in}\\[-0.1in]
\textbf{Bandwidth} & $\Delta\nu$ & 210 MHz \\
\hspace{0.2in}\\[-0.1in]
\textbf{Beam Width} & $\Theta$ & 2.5 deg\footnote{\cite{MacMahon:2009p6863}} \\
(per antenna, HPBW) & & \\
\hspace{0.2in}\\[-0.1in]
\textbf{Solid Angle} & $\Omega$ & 147.3 deg$^2$\footnote{Assuming all signal paths are operable but accounting for some dual-polarization observations, see Section
\ref{sec:inst}} \\
(Instantaneous) & & \\
\hspace{0.2in}\\[-0.1in]
\textbf{System Temperature}\footnote{Nominal value} & $T_{\rm sys}$ & 50 K\\
\hspace{0.2in}\\[-0.1in]
\textbf{Gain}$^{\rm c}$ & $G$ & $6.25 \times 10^{-3}$ K/Jy \\
\hspace{0.2in}\\[-0.1in]
\textbf{Dish Diameter} & $D$ & $\sim6$ m \\

\end{tabular}
\label{tab:feparam}
\end{minipage}
\end{table}

\section{Observations}
\label{sec:obs}
\begin{figure}[!ht]
\includegraphics[width=0.4\textwidth]{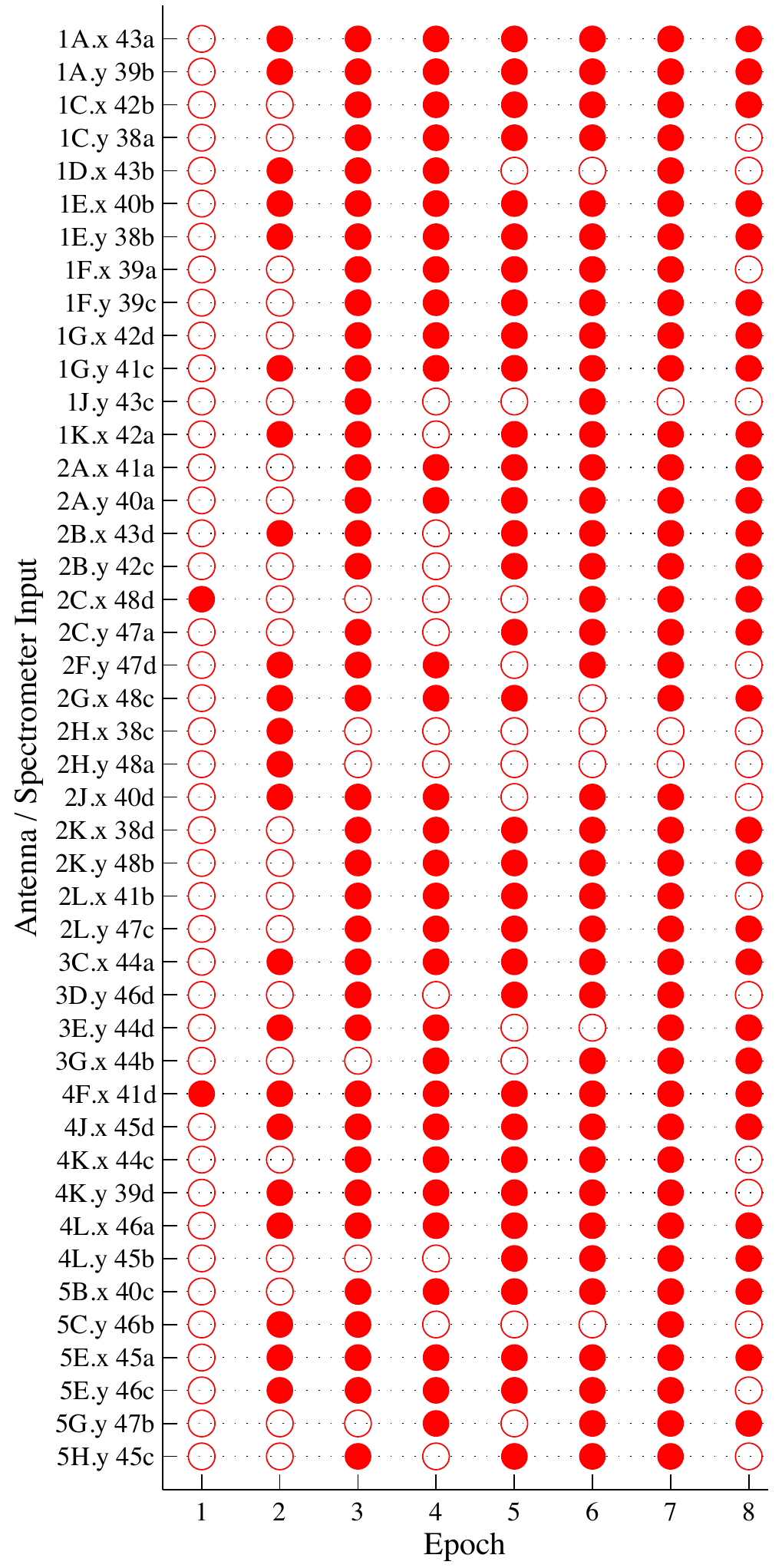}
\caption{
Diagram showing operability of each antenna/input as a function of epoch, based on detectability of B0329+54.  Filled circles indicate that a given signal path is operable.  For an input/epoch pair to be considered operable, B0329+54 must have been detected at every opportunity within the epoch. On the Y axis are each spectrometer input, labeled by their ATA antenna identifier followed by spectrometer input number.  The suffix on the antenna identifier indicates which of two dual linear polarization feeds was used.
\label{fig:operability_slim}
}
\end{figure}

During the period February 2008 to May 2008, we conducted approximately 480 hours of drift-scan observations with the Fly's Eye Spectrometer ( Table \ref{tab:feobs}).  
Data were collected in 60 minute intervals, each consisting of 58 minutes of drift observation followed by 2 minute diagnostic observations used for monitoring the health of the telescope and instrumentation.  Fly's Eye observations produced data at a rate of roughly 36 GB / hour, resulting in approximately 18 TB total collected data for the entire observation period.  The data are archived in Berkeley and available for analysis by request to the authors.  At the time of these observations, the ATA was undergoing commissioning, and a variety of system performance issues were being actively addressed.  The variation in SEFD from antenna to antenna and less-than-complete utilization of the 42 installed antennas are reflective of these issues.  To aid in dynamically determining signal path operability, a fixed pointing strip along a constant declination angle of $+54^{\circ}$, in which antennae were spaced 1 half-power beam width apart, was chosen for drift scan observations.  As the bright pulsar PSR B0329$+$54 drifted through the beam pattern at the sidereal rate, its detection or non-detection was used to determine whether or not a particular signal path was operable.  Declination $+54^{\circ}$ is well away from any significant source of galactic electron density confusion, the median maximum galactic DM contribution along this path is $\sim$68 pc cm$^{-3}$ (from the NE2001 model, \cite{Cordes:2002p2073}).  Figure \ref{fig:operability_slim} illustrates the overall observing efficiency after applying the B0329$+$54 detectability metric.  Out of a total of 921.1 input-days of observing, 579.9 input-days showed the expected detections of B0329$+$54, for a total observing efficiency of $\sim$63\%.  The poor efficiency in Epoch 1 likely reflects an unaccounted for change in our pointing script that directed antennas away from $+54^{\circ}$.  Although we believe most signal paths were operable, we have conservatively excluded these data.



\begin{table}[h!]
\begin{minipage}{3.8in}
\footnotesize
\caption{Fly's Eye Observations 02/2008$-$05/2008}
\centering
\vspace{0.1in}
\begin{tabular}{c c c}
\hline\hline\\[-0.10in]
\textbf{Epoch} & \textbf{MJD} & \textbf{Efficiency} \\
\hline
\hspace{0.2in}\\[-0.1in]
\textbf{1} & 54512.24 - 54515.97 & 5\% \\
\hspace{0.2in}\\[-0.1in]
\textbf{2} & 54519.20 - 54521.68 & 50\% \\
\hspace{0.2in}\\[-0.1in]
\textbf{3} & 54526.13 - 54528.64 & 86\% \\
\hspace{0.2in}\\[-0.1in]
\textbf{4} & 54533.04 - 54535.63 & 73\% \\
\hspace{0.2in}\\[-0.1in]
\textbf{5} & 54540.17 - 54542.59 & 75\% \\
\hspace{0.2in}\\[-0.1in]
\textbf{6} & 54547.18 - 54549.60 & 86\% \\
\hspace{0.2in}\\[-0.1in]
\textbf{7} & 54554.18 - 54556.62 & 93\% \\
\hspace{0.2in}\\[-0.1in]
\textbf{8} & 54561.26 - 54563.62 & 66\% \\
\end{tabular}
\label{tab:feobs}
\end{minipage}
\end{table}





\section{Analysis}
\label{sec:ana}
\subsection{Data Preparation}
Power spectra time series for each of the 44 inputs to the Fly's Eye Spectrometer were extracted as individual ``filterbank'' format files \citep{Lorimer:2000p10821}, broken into analysis chunks of length $2^{20}$ samples (representing 12 minutes).  This length was chosen to allow an entire analysis chunk and set of dedispersed time series to be kept in computer memory during analysis.  Prior to dedispersion, power spectra were normalized or ``equalized'' across both frequency channel and time.  Equalizing across frequency channels has the primary effect of correcting for the rippled bandpass imposed by both analog filter response and digital down conversion.  Equalization of accumulation values mitigates broadband gain changes and broadband impulsive interference.  This process was carried out as follows.  
  
We denote the power in channel $i$ at (discrete) time $t$ as $P_i(t) \in \left[0,255\right]$, and we computed a mean power per channel 
\begin{equation}
\overline{P_i} \equiv \frac{1}{T_0} \sum_{t=0}^{T_0-1} P_i(t)
\end{equation}
over some time period $T_0$.  $T_0$ is typically set to the length of an analysis chunk, 2$^{20}$ samples. A particular value $P_i(t)$ is then divided by the mean $\overline{P_i}$. i.e. the equalized value $P_i'(t) \equiv P_i(t) / \overline{P_i}$. The mean power in each channel is then unity, since
\begin{equation}
\frac{1}{T_0} \sum_{t=0}^{T_0-1} P_i'(t) = 1
\end{equation}
Mean power equalization was performed on the frequency spectrum equalized values $P_i'(t)$.  The power mean over all frequency channels for a single integration (time sample $t$) is defined as 
\begin{equation}
\overline{P'(t)} \equiv \frac{1}{N} \sum_{i=0}^{N-1} P_i'(t)
\end{equation}
$N$ is the number of channels.  With the mean powers $\overline{P'(t)}$, we can define the equalization of the powers $P_i'(t)$. The mean power equalized values $P_i''(t) \equiv P_i'(t) / \overline{P'(t)}$. This procedure ensures that the sum of the power samples for any time $T_0$ is normalized such that $\sum_{i=0}^{N-1} P_i''(t) = N$, effectively flattening the DM = 0 time series.

\begin{figure}[ht!]
\includegraphics[width=0.45\textwidth]{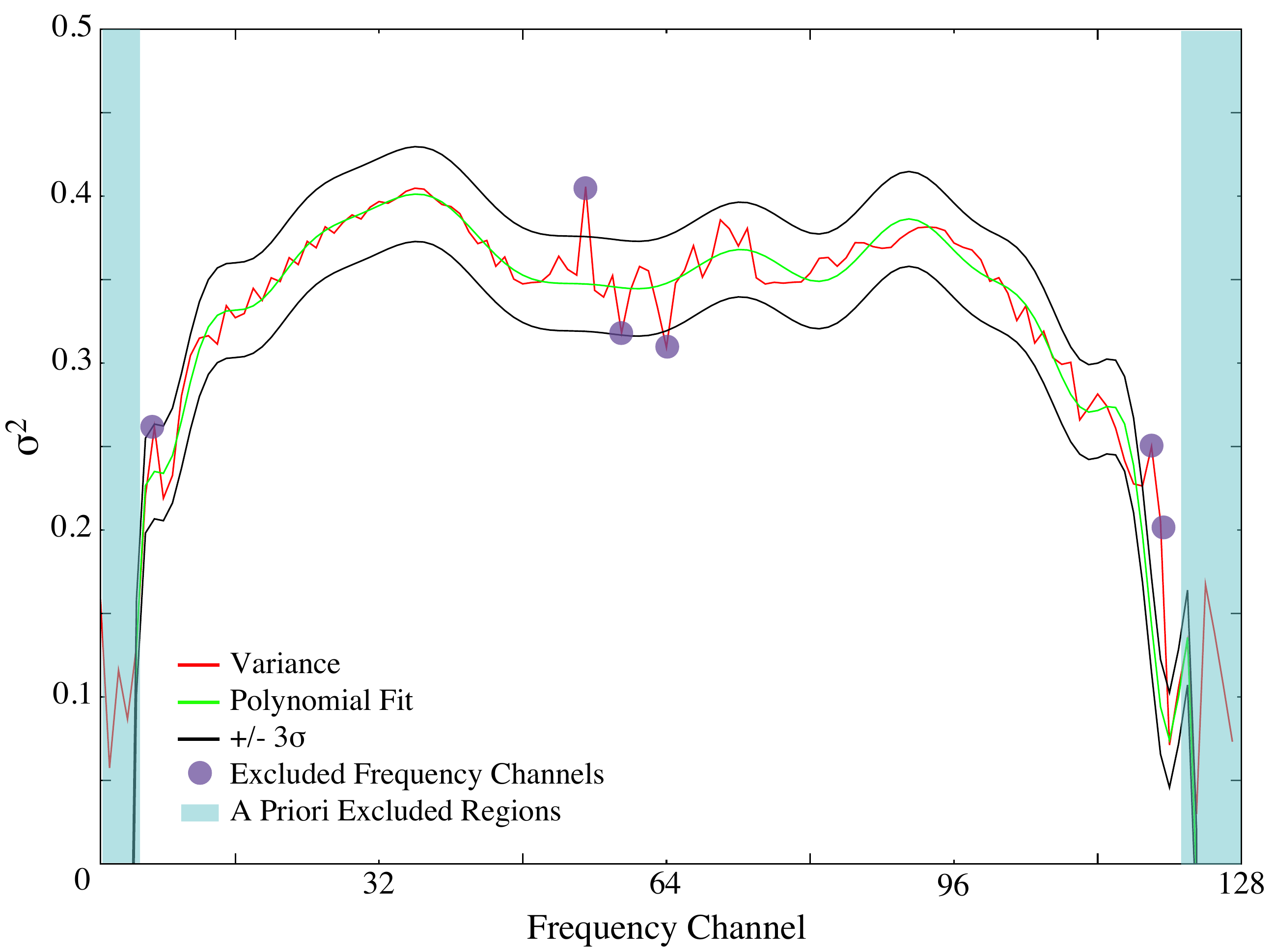}
\caption{
Variance vs. frequency channel for a 2$^{20}$ spectra Fly's Eye observation (red), an iterative polynomial fit to the variance curve (green) and $\pm3\sigma$ bounds on the fit (black).  Band edges are excluded a priori from the polynomial fit, as instrumental response is poor in these regions due to filter roll-off.  Band edges and individual frequency channels with high variance are excluded from the de-dispersion process. \label{fig:variance}
}
\end{figure}

Prior to the equalization process, individual frequency channels with especially large amounts of interference were identified and logged.  Our algorithm used the variance of each frequency channel over an analysis chunk as a measure of the amount of interference in that channel.  Using the previously defined quantities, we computed the variance of the values $P_i(t)$ for each channel $i$ over $T_0$, and then fit a polynomial to the resulting curve ${\rm{Var}}(P_i)$ (Figure \ref{fig:variance}).  Frequency channels for which the computed variance differed from the polynomial fit by ${\rm{Var}}(P_i)>2\sigma$ were excluded from subsequent dedispersion.  We explicitly excluded 8 frequency channels at the top of the band and 13 frequency channels at the bottom of the band due to analog filter roll off and the presence of bright air route surveillance radar below 1350 MHz.
 
\subsection{Single Pulse Search}
A single pulse search in dedispersed time series for events having a signal-to-noise ratio (SNR) greater than five standard deviations above the mean, $\sigma > 5.0$,  was carried out over 744 trial DMs between 0-2002 pc cm$^{-3}$ using the SigProc tools \citep{Lorimer:2000p10821}.  An approximate matched filtering algorithm was employed to increase sensitivity to broadened pulses in which each dedispersed time series was iteratively smoothed by adding 2$^{n}$ adjacent time samples over the range n = 0 to 10, following \cite{Cordes:2003p7163}.  This results in an effective box car smoothing of maximal window size 2$^{10}$ samples or 0.64 to 5.12 s, depending on the level of time collapse (Table \ref{tab:dmsearch}).  The initial DM step size was set such that the time delay associated with the DM step was less than the sampling interval.  Spectra were iteratively collapsed in time by a factor of two at DMs 329.0, 658.0, 1314.0 pc cm$^{-3}$ to speed analysis, as detailed Table \ref{tab:dmsearch}.  At these thresholds a pulse would subtend a minimum of 2, 4 or 8 spectra at the top of the band, respectively, and thus halving the effective time resolution imposes no loss in sensitivity.  Similarly, the DM step size can be reduced by the same factor and remain less than the new effective time resolution.  These search parameter choices were all designed to ensure that the dominant source of temporal smearing was due to the  unavoidable (at the time of analysis) integration time smearing and in-band dispersive smearing.  Like many problems in radio astronomy, the analysis of multibeam pulse search data is readily parallelizable.  Here we distributed 58 minute observations to individual compute nodes, and parallelized inputs over individual CPU cores.  In total, our analysis consumed $\sim$20,000 CPU-hours.

\begin{table}[h!]
\begin{minipage}{3.8in}
\footnotesize
\caption{Single Pulse Search Parameters}
\centering
\vspace{0.1in}
\begin{tabular}{l c l c c}
\hline\hline\\[-0.10in]
\multicolumn{3}{c}{\textbf{Dispersion Measure Range}}& \textbf{$\Delta$DM} & \textbf{Time Collapse Factor} \\
\hline
\hspace{0.2in}\\[-0.1in]
0.0 & - & 329.0 pc cm$^{-3}$& 1.0 & 1 \\
\hspace{0.2in}\\[-0.1in]
330.0 & - & 656.0 & 2.0 & 2 \\
\hspace{0.2in}\\[-0.1in]
658.0 & - & 1310.0& 4.0 & 4 \\
\hspace{0.2in}\\[-0.1in]
1314.0 & - & 2002.0& 8.0 & 8 \\
\end{tabular}
\label{tab:dmsearch}
\end{minipage}
\end{table}

\subsection{Post Processing RFI Filter}




After dedispersion, any strong signal present in a dynamic spectra -- be it from interference or a real event -- will be detected at multiple DMs, strengths and times.  The distribution of these detections depends on the observed properties of the signal, the range of DMs searched and any pre- or post-dispersion processing applied.  In the case of quadratically chirped radio pulses, this distribution follows a characteristic functional form \citep{Cordes:2003p7163}.  Likewise, certain kinds of interference will exhibit predictable detection distributions.  Wideband, temporally narrow RFI will be detected over many DMs with the highest strength detections at low DMs.  Narrowband, long duration RFI will also be detected at many DMs but will peak in strength when the dispersion path for a trial dispersion measure optimally overlaps the narrow-band interference.  Finally, wideband and long duration interference or rapid gain changes will cause an excess of detections in all DMs for the duration of the event.

\begin{figure}
    \centering
    \subfloat[]{\label{crab_rfi:p0}\includegraphics[scale=0.31]{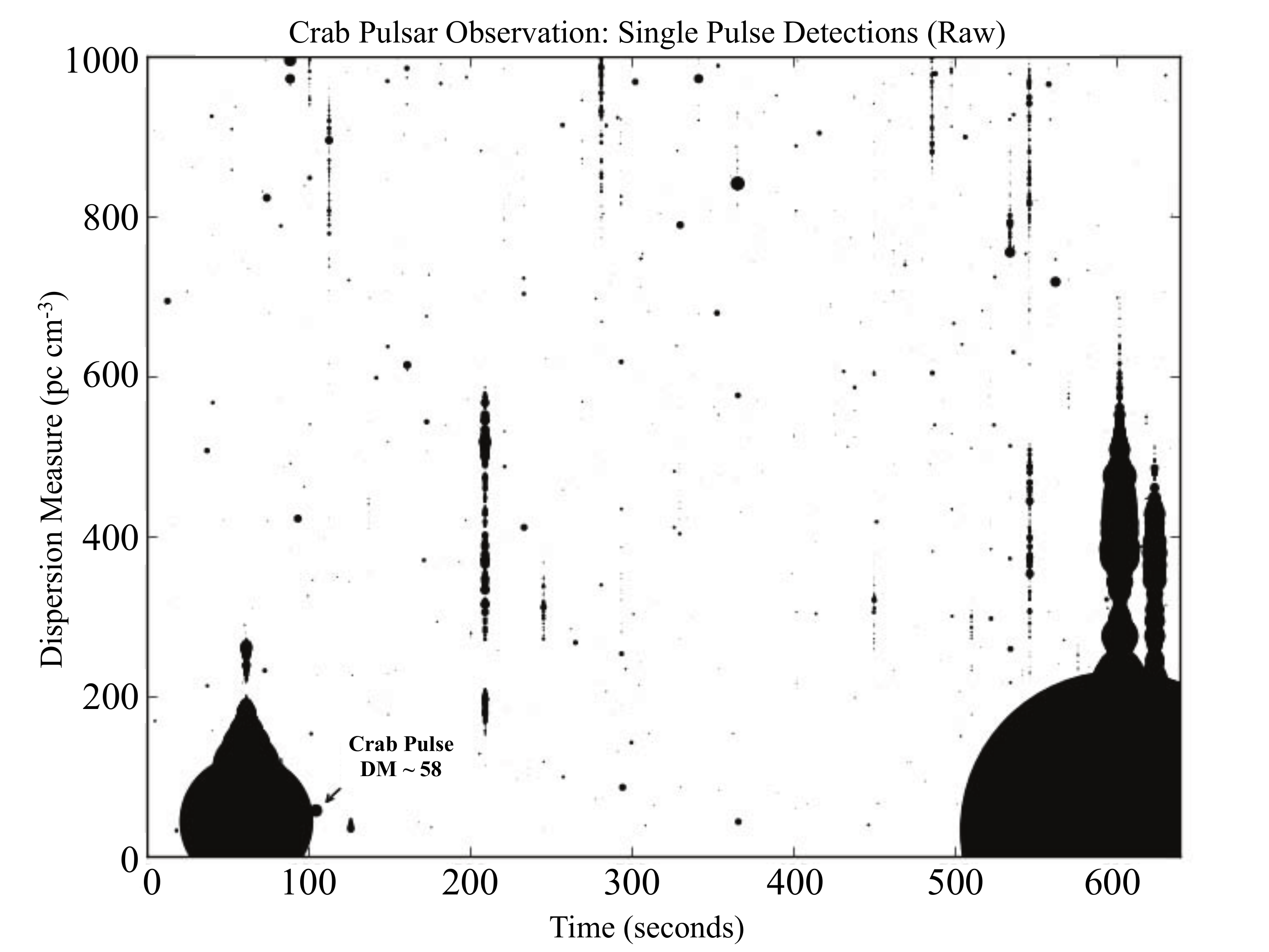}}
    \subfloat[]{\label{crab_rfi:p2}\includegraphics[scale=0.31]{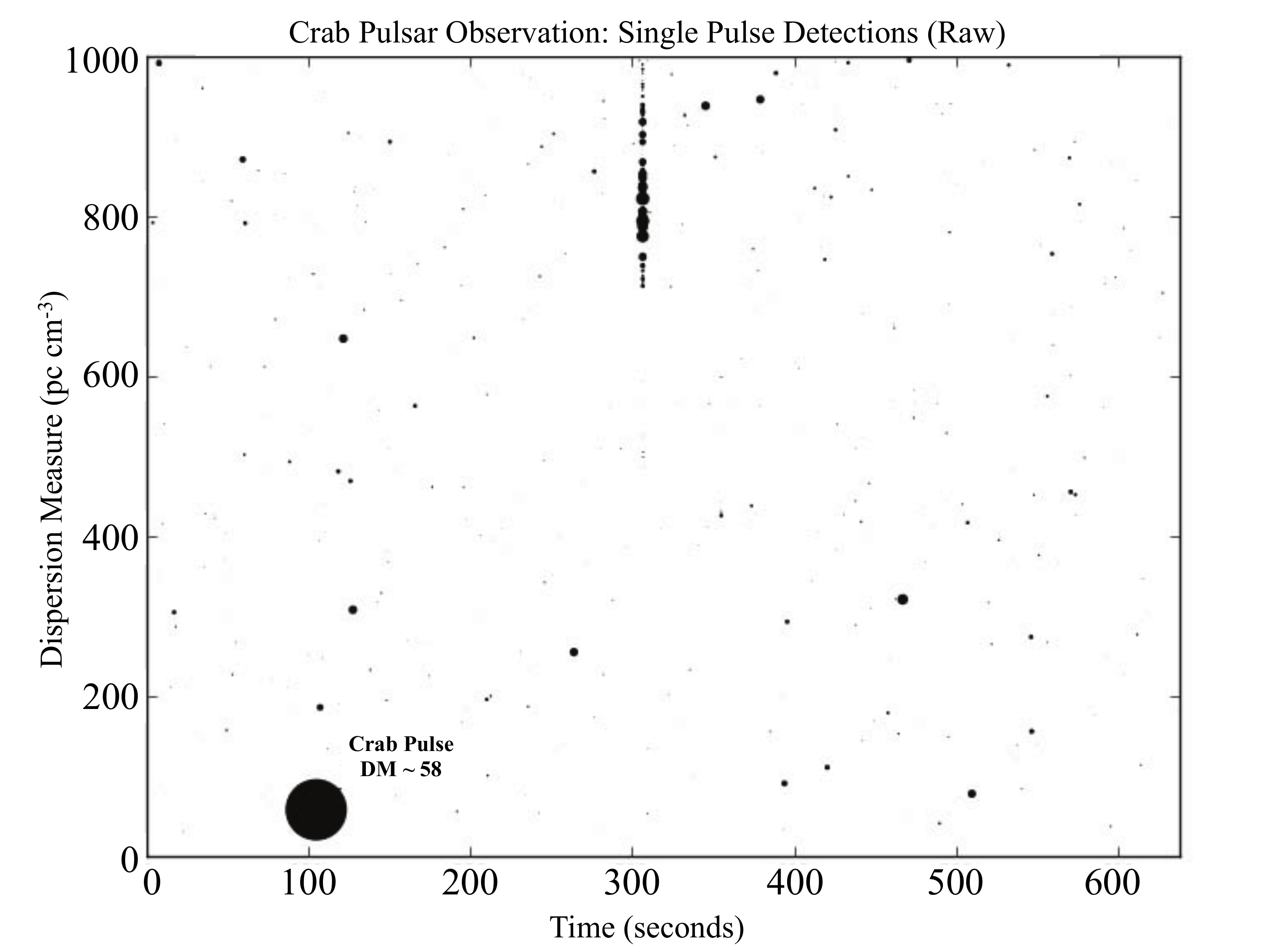}}\\
    \subfloat[]{\label{crab_rfi:p1}\includegraphics[scale=0.31]{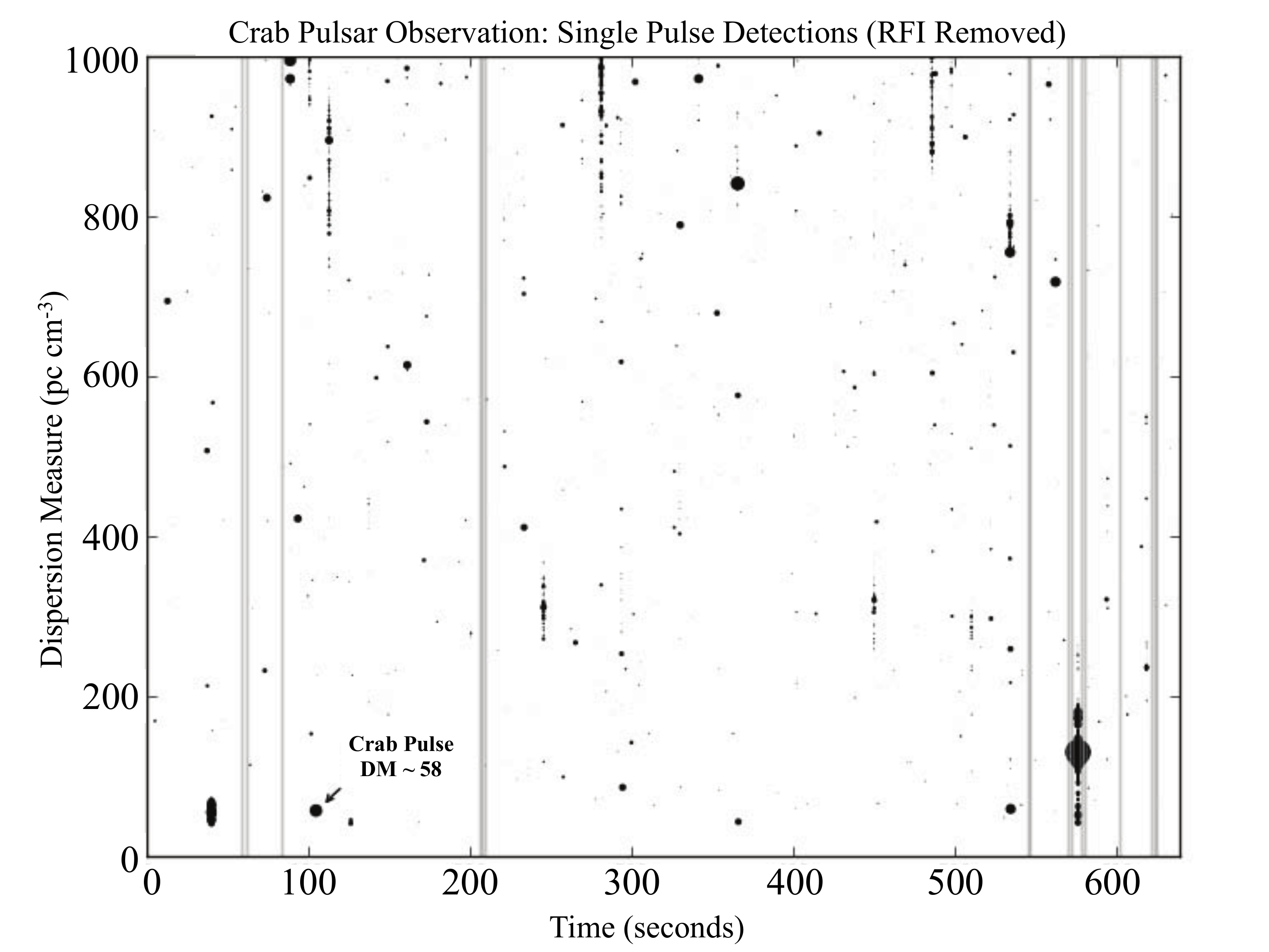}}
    \subfloat[]{\label{crab_rfi:p3}\includegraphics[scale=0.31]{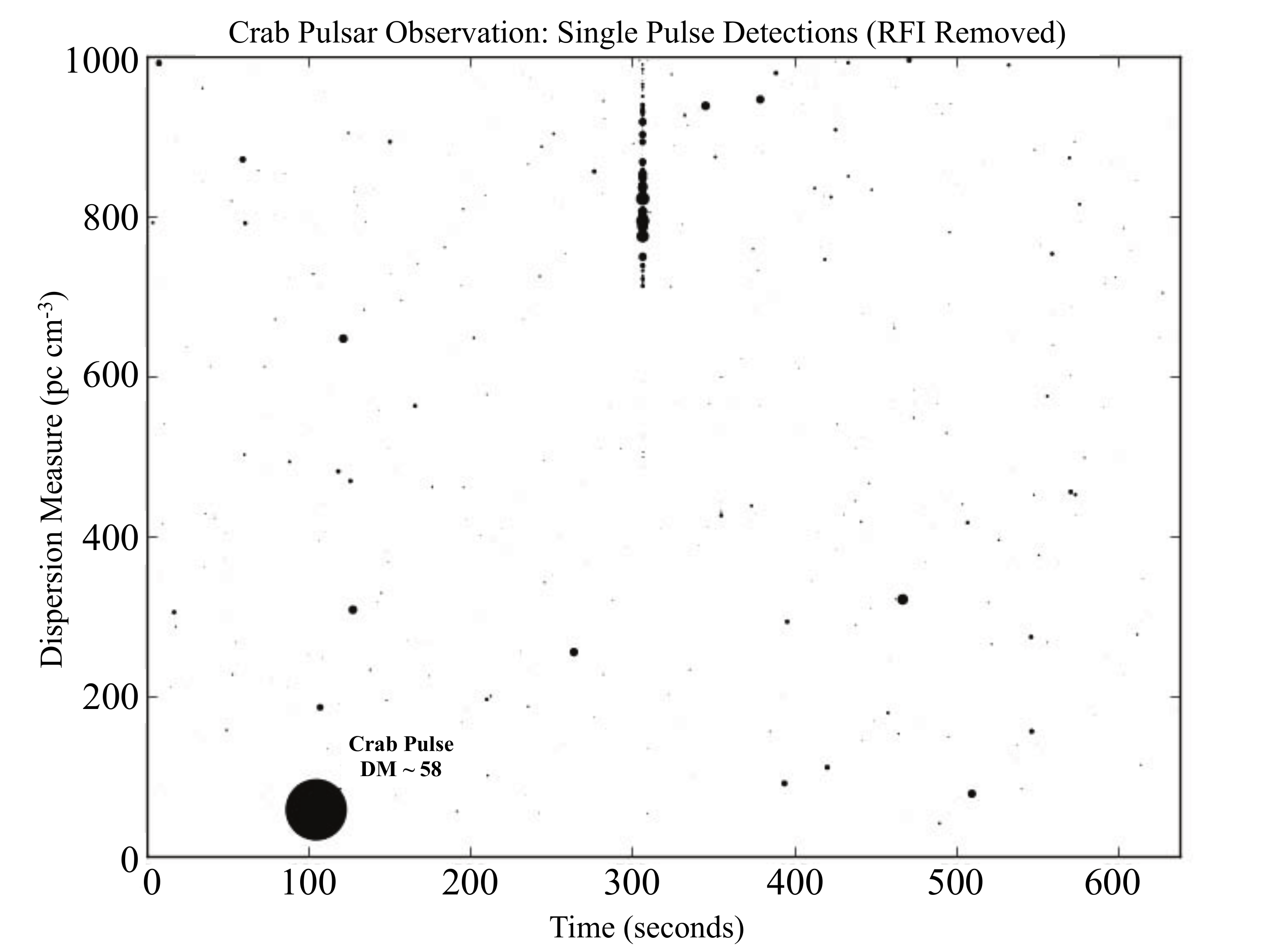}}
    \caption{Detection of giant pulses from the Crab pulsar for two individual FE inputs before (\ref{crab_rfi:p0}, \ref{crab_rfi:p2}) and after (\ref{crab_rfi:p1}, \ref{crab_rfi:p3}) application of a post processing RFI filter.  Each plot shows events vs. time and trial dispersion measure, here the radii of plot points are proportional to $({\rm{signal\ to\ noise}})^2$}
    \label{fig:crab_rfi}
\end{figure}

As a first cut on the vast number of high SNR candidates detected, we flagged any 1 second input-time region where the highest SNR candidate in that region was $< 50$ pc cm$^{-3}$ or  $> 1950 $ pc cm$^{-3}$, or the total number of pulses over 5$\sigma$ in the same DM regimes exceeded a factor of 4 times the mean number of pulses in each regime.  The DM $< 50$ pc cm$^{-3}$ would reject any relatively nearby galactic events, but distinguishing true astrophysical bursts from interference at these low DMs is very difficult because of the correspondingly small quadratic chirp.  Further, our focus was primarily on potentially extragalactic events for which the DM contribution from the Milky Way and host ISM should well exceed this threshold.  Figure \ref{fig:crab_rfi} shows the results of applying this metric to pulse detections from two observations of the Crab Pulsar.  The algorithm was effective at rejecting strong interference and avoided rejecting true astrophysical events.  Strong interference dominates Figure \ref{crab_rfi:p0}, but is greatly diminished in \ref{crab_rfi:p0}.  Figures \ref{crab_rfi:p2} and \ref{crab_rfi:p3} shows the detection of a bright pulse left untouched after applying our algorithm.

\begin{figure}[htb]
\centering\includegraphics[width=0.8\textwidth]{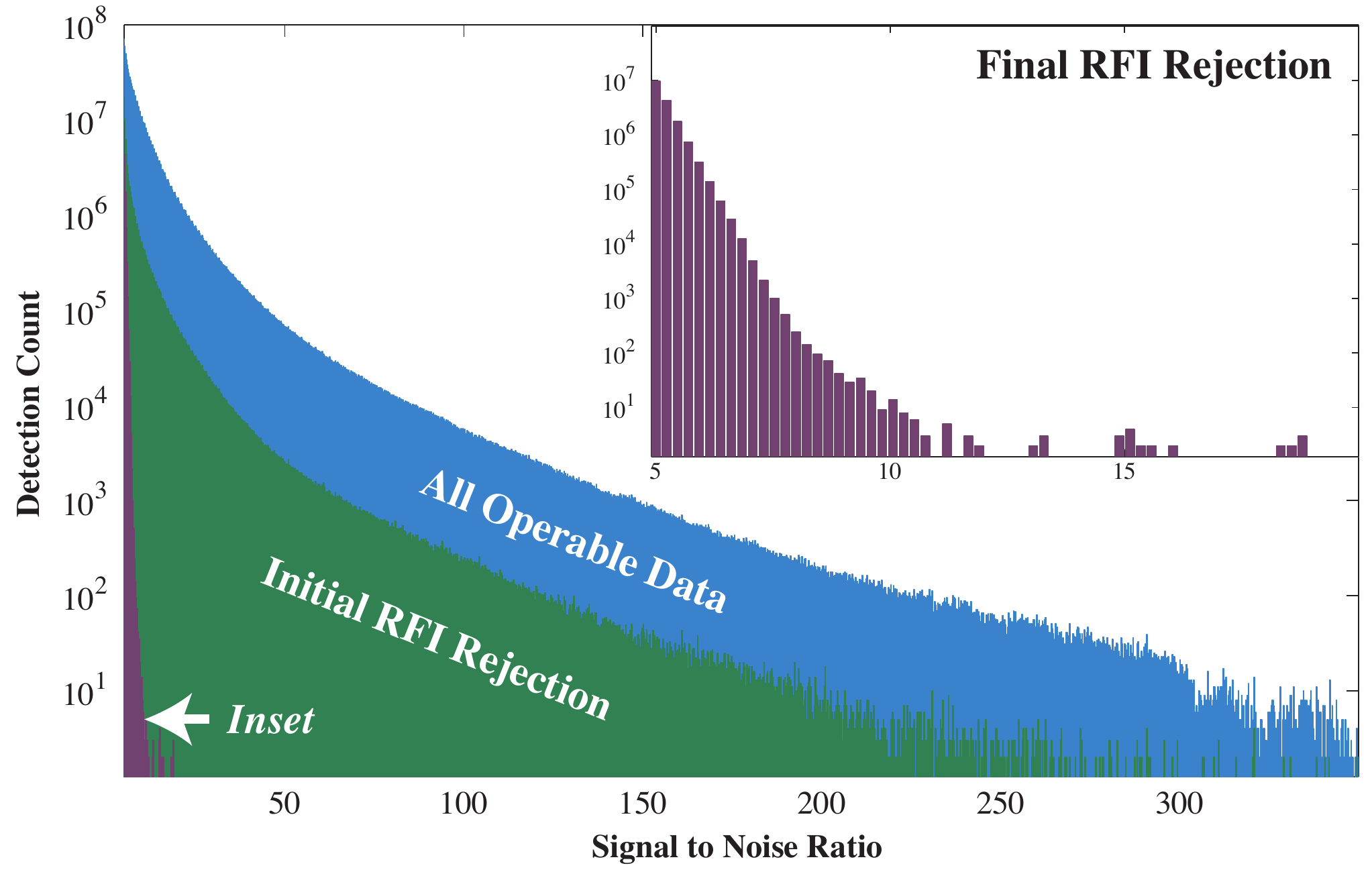}
\caption{SNR histograms for all pulses detected in operable signal paths, all pulses in operable paths after applying the first cut RFI rejection, and (inset) all pulses detected in $2^{20}$ sample observation chunks having a mean pulse detection SNR $\sigma < 5.5$.
\label{fig:fehist}
}
\end{figure}

\begin{figure}[htb]
\centering\includegraphics[width=0.8\textwidth]{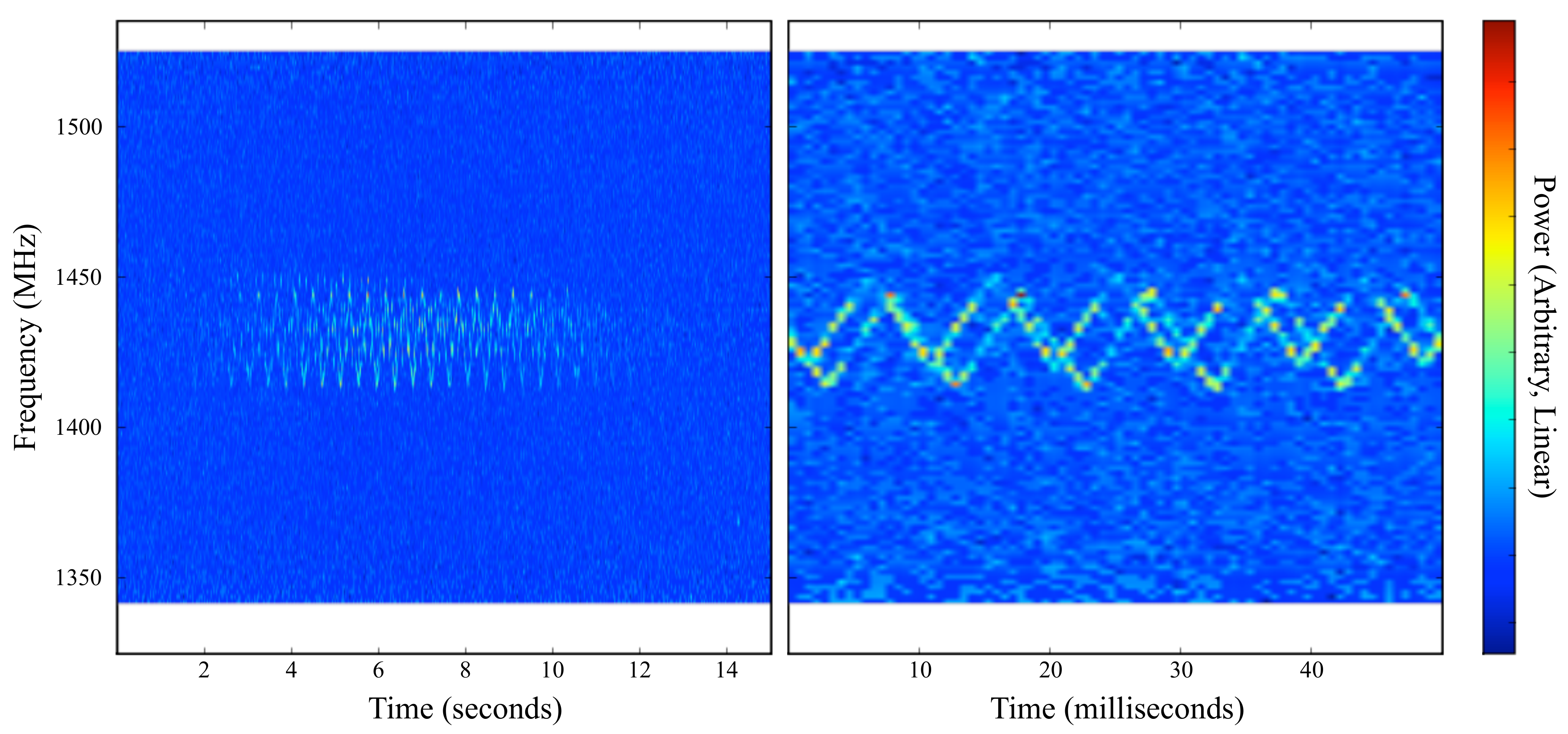}
\caption{A spectrogram plot of intensity in the time-frequency plane of a satellite interferer that passed our RFI rejection algorithms.  Edges of the observed band, where system response is poor, have been masked.
\label{fig:satellite_detection}
}
\end{figure}

Following RFI rejection, all events with a SNR $\sigma > 8.0$ detected in $2^{20}$ sample analysis chunks for which the mean SNR of all detections was $\overline{\sigma} < 5.5$ were selected for visual analysis.  Plots similar to Figure \ref{fig:crab_rfi} were examined for each detection.  If an event did not appear to be interference, we extracted and closely examined $t$ vs. $\nu$ spectrograms of 1, 2, 4 and 10 seconds around the event.  Figure \ref{fig:fehist} shows the SNR distributions for all pulses detected in operable signal paths, all pulses in operable paths after applying the first cut RFI rejection algorithm described above, and (inset) all pulses detected in observations having a mean pulse detection SNR $\sigma < 5.5$.  Upon close inspection, none of the pulse candidates identified appeared to be of astronomical origin.  An example of the pathological interference that escaped our interference rejection algorithms is shown in Figure \ref{fig:satellite_detection}.  Strong pulses were detected in regions where dedispersion curves aligned with the triangle-wave modulation of the interferer, at a DM of $\sim$80 pc cm$^{-3}$.  This particular interferer was detected at multiple epochs, and appears to originate with an orbiting satellite.

\section{Discussion}
\label{sec:discussion}
\begin{figure}[htb]
\centering\includegraphics[width=0.8\textwidth]{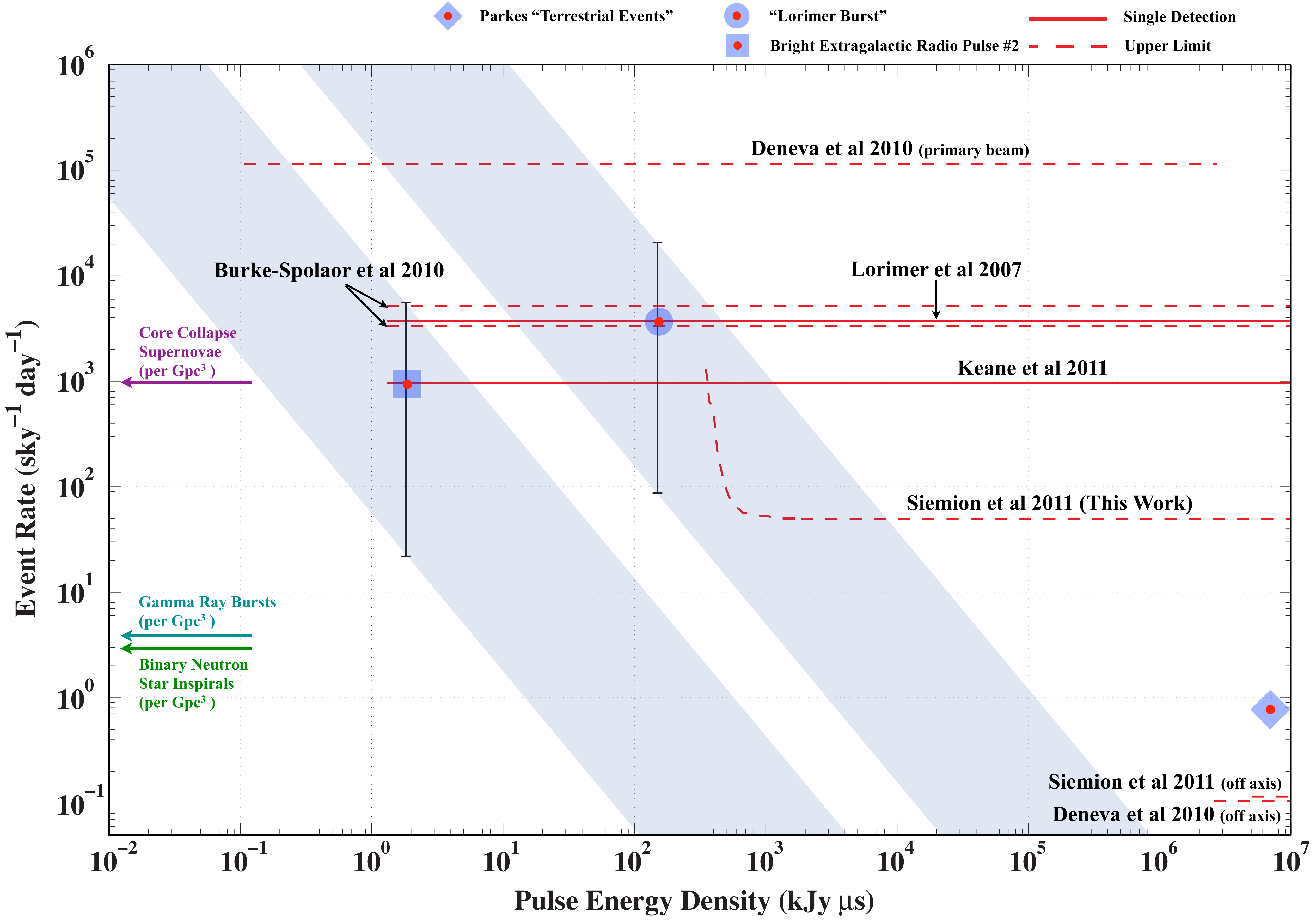}
\caption{Pulse energy density vs. rate limit for the surveys in Table \ref{tab:pulsesurveys}.  Rate limit curves assume a 10 millisecond pulse duration.  Shaded bars on the Lorimer Burst and the extragalactic event described in \cite{Keane:2011p10822} represent 2 sigma confidence \citep{Gehrels:1986p10835} on the Euclidean isotropic distribution.  The point for terrestrial events identified in \cite{BurkeSpolaor:2010p5814} assumes off-axis detection.  Rates of core collapse supernovae, gamma ray bursts and binary neutron star inspirals in a 1 Gpc$^{3}$ volume are taken from \cite{Madau:1998p10935}, \cite{Guetta:2007p11057} and \cite{Kalogera:2004p11110}, respectively, via \cite{Lorimer:2007p5652}.  Here we assume flat sensitivity across the HPBW of each receiver beam and do not take into account reduced sensitivity in wide field-of-view sidelobes, except in the case of this work and \cite{Deneva:2009p4149}.  The curved line for the Siemion et al survey reflects variation in antenna system temperature.
\label{fig:ratelimit}
}
\end{figure}

Our results indicate that the millisecond radio sky is relatively quiescent at the energies probed by our experiment.  For a threshold of $8\sigma$, our minimum detectable energy density is $E_{\rm{min}}= 111$ kJy $\mu$s or a mean flux density of $\sim$178 Jy for a 625 $\mu$s pulse.  Based on our non-detection, pulses of these energy densities originating with an isotropic progenitor must occur at a rate less than 2 sky$^{-1}$ day$^{-1}$.   We can determine the limiting rate of occurrence of bursts as a function of antenna sensitivity by computing the rate 
\begin{equation}
\eta(E > E_{\rm{min}}) = \frac{{1}}{{\sum {\Omega \cdot T_{{\rm{obs }}}(SEFD > SEFD_{\rm{min}}) } }}
\end{equation}
where $E_{\rm{min}}$ corresponds to the detectable limit for pulses of a given duration in an antenna having $SEFD_{\rm{min}}$.  Figure  \ref{fig:ratelimit} shows the event rate limit calculated from these results alongside other recent single pulse searches in Table \ref{tab:pulsesurveys}.  Our rate limit does not yet sample potential coherent radio emission processes from neutron star binary inspirals or gamma ray bursts, but we can place an order of magnitude limit on the luminosity of coherent radio emission from core collapse supernovae (CC SNe).  Assuming isotropic emission, our survey could have detected $\sim$20 10 ms events with an intrinsic rate of 1000 sky$^{-1}$ day$^{-1}$ at an apparent energy density limit of $E_{\rm{limit}}=10^3$ kJy $\mu$s.  Assuming that coherent radio emission from CC SNe is beamed over a solid angle $\Omega$, we can limit the emission cone to $\Omega<4\pi/20$.  Translating this into an upper limit on luminosity using the radius of a spherical volume of 1 Gpc$^3$ , we have 
\begin{equation}
L < E_{\rm{limit}} \Delta\nu \frac{{4 \pi R_{\rm{max}}^2}}{{20}} \frac{{1}}{{\Delta t}}  
\end{equation}
For emission over a bandwidth $\Delta\nu = 1$ GHz and $\Delta t = 10$ ms we have L $<$ $\sim2\times 10^{42}$ erg sec$^{-1}$.

As discussed in \cite{Lorimer:2007p5652} and \cite{Keane:2011p10822}, if the two detected pulses inferred to be at cosmological distances are indeed real, they must represent an entirely new source class.  The extreme SNR of the two sparse detections is curious, and as previously discussed by several authors, contradicts the assumption of an isotropically distributed population.  In the case of this experiment, we would have expected to detect $\sim$15 events similar to the \cite{Lorimer:2007p5652} burst, but $\ll1$ similar to the \cite{Keane:2011p10822} extragalactic event.  \footnote{Assuming an Euclidean isotropic distribution and correcting for the Fly's Eye survey's lower sensitivity.  See, for example, \cite{Deneva:2009p4149} for an exposition of this calculation.} The incompatibility of the implied rates for the two isolated detections is difficult to explain.  The galactic latitude of the two events differ significantly, $b=-41.8^{\circ}$ for the \cite{Lorimer:2007p5652} event and $b=-4^{\circ}$ for the \cite{Keane:2011p10822} burst.  While one might guess that being much closer to the galactic plane would make it easier to explain a large dispersion measure, of the 227 known pulsars between 3$^\circ$ and 5$^{\circ}$ off the galactic plane, none has a DM $>$ 460 pc cm$^{-3}$, standing in stark contrast with the \cite{Keane:2011p10822} event at a DM of 745 pc cm$^{-3}$    However, the increased intragalactic path length of this detection does present more opportunity for an unseen highly ionized nebula to make an aberrant contribution to the total integrated free electron column density.        

The 15 \cite{BurkeSpolaor:2010p5814} detections also remain puzzling.  Assuming the phenomena that generated these bursts is not unique to the Parkes site, we can estimate the number of similar events that the Fly's Eye survey should have detected.   Assuming that these bright events were observed in far out side lobes, a reasonable comparison between the two surveys reduces to simply the ratio of their observing time, thus the Fly's Eye survey should have detected 

\begin{equation}
15 {\rm{\ events}} \times \frac{{450}}{{346.1+532.4 }}\rm{h}\sim 10 {\rm{\ events}}
\end{equation}
of these events as well.  Note that here we assume antenna efficiency and system temperature differences are negligible.  We will refrain from speculating on the cause of these events, but note that our observations well sampled the diurnal cycle as well as varying levels of precipitation.  We are unaware of any lightning activity in the near vicinity of the ATA during our observations. 

It remains perplexing as to why all of these unique transient bursts have been detected only at the Parkes telescope.  However, with the aggressiveness with which interference must be excised in such experiments, and the varying means by which it is accomplished, we speculate that it is not wholly out of the question that some processing pipelines could be better tuned to detecting single pulses at extragalactic DMs or other unexpected characteristics.  The fact that all of these pulses were themselves discovered in reanalyses of previously mined surveys, with the second extragalactic event discovered in a 3rd reanalysis, indicates that other extant surveys may harbor additional as-yet undetected events.  Our own pipeline has been honed by the exercise of this experiment, and a future search would undoubtedly be superior.  For instance, summing polarizations would yield an additional $\sqrt 2$ sensitivity on some antennas.  We attempted multi-beam coincidence RFI rejection, excluding pulses found at similar DMs and similar times in multiple antennas, and found it somewhat ineffective.  We attribute the poor performance of this algorithm to a number of factors, among them that our algorithm didn't account for variations in signal path sensitivity and multi-path propagation of interferers led to strong variation in detection signal-to-noise from antenna to antenna. 

 In previous searches where multi-beam excision has been effective, multiple receivers were co-located in the focal plane of a single telescope.  Our results indicate that the interference environment is generally better correlated in these more confined configurations.  Even so, applying a multi-beam coincidence metric in the time/frequency plane, prior to baseline subtraction and thresholding, would undoubtedly perform better.  We explored these possibilities in parallel with the analysis described here, in anticipation of a future reanalysis, in \cite{Hogden:2011p12088}.  Of the methods explored on a subset of Fly's Eye data in \cite{Hogden:2011p12088}, we find that a combination of Huber filtering and adaptive interference cancellation performs optimally for the types of interference seen in this experiment.  Friends-of-friends algorithms, as described in \cite{Deneva:2009p4149} or Hough transform-based detection \citep{Fridman:2010p11233} would also improve our ability to detect low SNR events.  In any case, additional surveys and analyses will soon detect more examples of isolated radio pulses which will expand the available sample.  New wide-field interferometric radio telescopes will be well equipped to not only detect additional events, but also provide much better localization than has so far been possible with single dish facilities.  Precise localization would be extremely useful in differentiating between true extragalactic events and those that may appear so because of passage through highly ionized intragalactic regions.  

\section{Summary}

We have developed a novel multiple-input digital spectrometer which we have used to conduct a wide field search for bright dispersed radio pulses using the Allen Telescope Array.  This wide field search yielded no detections, allowing us to place a limiting rate of less than 2 sky$^{-1}$ hour$^{-1}$ for 10 millisecond duration pulses having mean apparent flux densities greater than 44 Jy.  The flux densities probed by this experiment are well above individual pulses from known pulsars and RRATs, just grazing the very brightest of the giant pulse producing pulsars, none of which are present in the field surveyed.  We have placed new limits on very bright coherent emission from events similar to the singular event described in \cite{Lorimer:2007p5652}.  Our results indicate that the  \cite{Lorimer:2007p5652} event must belong to a very rare source class, if it is indeed astrophysical.  We did not detect any quadratically dispersed terrestrial interference similar to that seen at the Parkes observatory, e.g. \cite{BurkeSpolaor:2010p5814}, consistent with other non-Parkes surveys, e.g. \cite{Deneva:2009p4149}.

The work presented here has shown that sources of bright fast transient radio emission must be relatively rare.  We have also shown both the utility, and associated of challenges, of using an interferometric array in a multi-pointing ``fly's eye'' mode.  The next generation of radio interferometers currently being built in preparation for the Square Kilometer Array \citep{Dewdney:2009bq} will offer new opportunities to explore the fast transient regime with greater sensitivity over large solid angles.  The use of these new instruments for fly's eye mode surveys will require making individual antenna or station data accessible to sufficient digital hardware, and we encourage this consideration to be taken into account during the design phase.  Commensal surveys for fast transient emission with interferometers, using the incoherent sum of antennas pointed in the same direction, could offer an excellent trade-off between sensitivity and solid angle while incurring little additional hardware cost and no additional observing time, see e.g. \cite{Macquart:2011p11135}.  Again, such capabilities will require consideration early in the design process in order to be realized efficiently.  The exploration of the fast radio transient parameter space is just beginning, and the contrasting results of this and other experiments clearly indicate we have much yet to learn.


\section{Acknowledgements}
The ATA is jointly operated by the University of California, Berkeley Radio Astronomy Lab and the SETI Institute in Mountain View, California.  The authors would like to acknowledge the generous support of the Paul
G. Allen Family Foundation, who have provided major support for design, construction,
and operations of the ATA. Contributions from Nathan Myhrvold, Xilinx Corporation, Sun
Microsystems, and other private donors have been instrumental in supporting the ATA.
The ATA has been supported by contributions from the US Naval Observatory in addition
to National Science Foundation grants AST-050690 and AST-0838268.  

We thank Scott Ransom, Matthew Bailes, Eric Korpela and Josh von Korff for productive discussions and Garrett Keating for providing the interferometric standard calibrator observations from which our sensitivity estimates are derived.  We also thank the Collaboration for Astronomy Signal Processing and Electronics Research and its members for contributing to the instrumentation development framework used in construction of the Fly's Eye spectrometer.
We thank Henry Chen, Jeff Cobb, Matt Dexter, Terry Filiba, Rick Forster, Colby Gutierrez-Kraybill, Matt Lebofsky, David MacMahon and Jason Manley for invaluable engineering and data management support and expertise.

We also thank an anonymous reviewer for a careful reading of an earlier draft of this manuscript and thoughtful comments which have improved this work.




\begin{figure}[htb]
\includegraphics[width=0.8\textwidth]{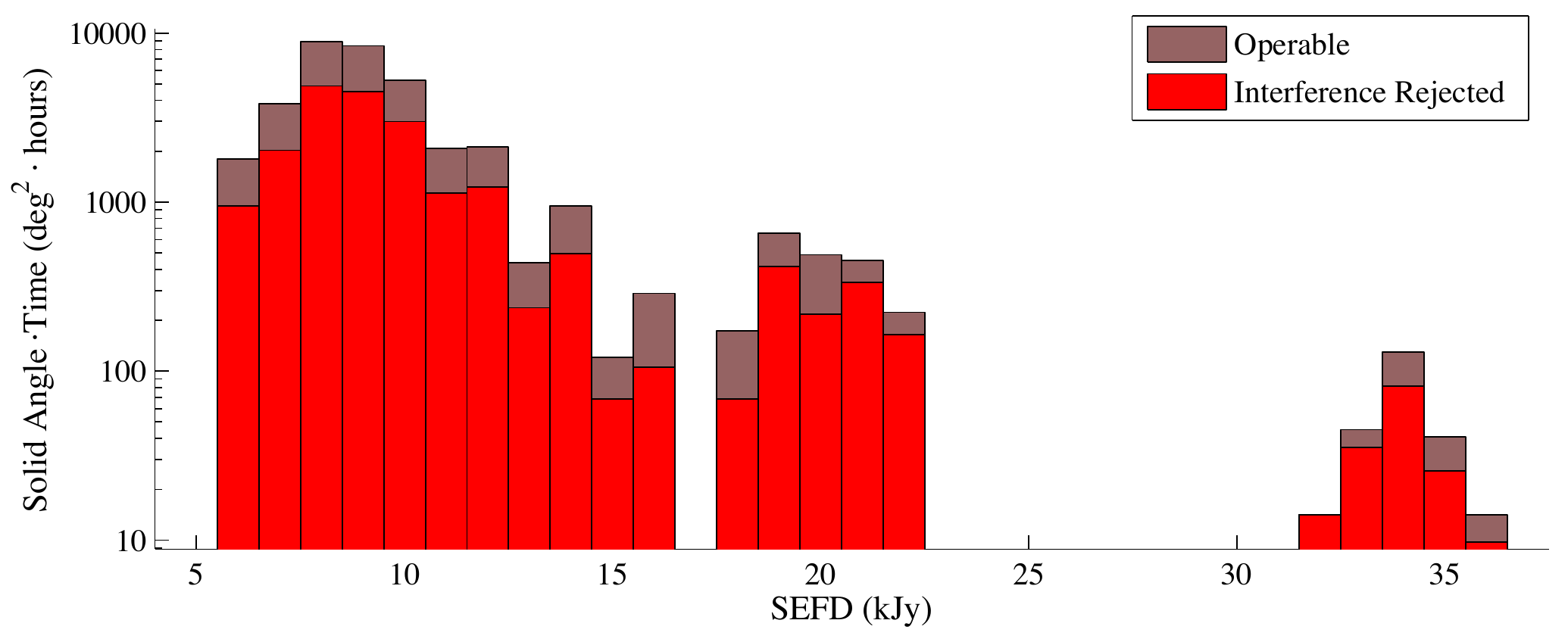}
\caption{
A plot of system sensitivity, described by the system equivalent flux density, vs. total operable observing, described by the observing solid angle $\cdot$ time product.  Plot includes both operable observing, taken from the set of signal paths where PSR B0329$+$54 was consistently detected, and the final low-RFI analysis set.  In cases where two polarizations were observed for a given antenna, the lower SEFD was used.
\label{fig:cover_strip}
}
\end{figure}



\begin{table}
\begin{minipage}{5.8in}
\footnotesize
\caption{Recent L-band Single Pulse Surveys}
\centering
\vspace{0.1in}
\begin{tabular}{l c c c}
\hline\hline\\[-0.10in]
\textbf{Survey} & \textbf{T$_{{\rm{obs}}}$} & \textbf{Total Solid Angle} &\textbf{E$_{\textrm{min}}$}\footnote{For a 10 millisecond pulse.  Values given here are conservative, see references for details.} \\
 & (hours) & (deg$^2$) & kJy $\mu$s\\
\hline
\hspace{0.2in}\\[-0.1in]
\cite{Edwards:2001p6386}\\ Re-analysis by \cite{BurkeSpolaor:2010p6170} & 346.1 & 0.556\footnote{Parkes Multibeam - 14$'$ HPBW/beam $\times$ 13 beams} & 0.8\\ 
\hspace{0.2in}\\[-0.1in]
\cite{Manchester:2006p12002}, \\ Re-analysis by \cite{Lorimer:2007p5652} & 480.7 & 0.556 & 0.8\\ 
\hspace{0.2in}\\[-0.1in]
\cite{Jacoby:2009p6050},\\ Re-analysis by \cite{BurkeSpolaor:2010p6170} & 532.4 &  0.556  & 0.8 \\ 
\hspace{0.2in}\\[-0.1in]
\cite{Manchester:2001p6299}, \\Re-analysis by \cite{Keane:2010p5855} and \cite{Keane:2011p10822} & 1864.3 & 0.556 & 0.8\\ 
\hspace{0.2in}\\[-0.1in]
\cite{Deneva:2009p4149} & 461.0 & 0.0187 \footnote{Arecibo ALFA - 3.5$'$ HPBW/beam $\times$ 7 beams} & 0.1 \\
\hspace{0.2in}\\[-0.1in]
Fly's Eye (this work) & 136 & 147 \footnote{2.5$^{\circ}$ HPBW/beam $\times$ 30 beams} & 440 \footnote{Using a nominal SEFD of 8 kJy}\\
\end{tabular}
\label{tab:pulsesurveys}
\end{minipage}
\end{table}

\clearpage

\bibliography{references}
\appendix

\clearpage

\end{document}